\documentstyle[prl,aps,twocolumn,floats,epsf]{revtex}

\newcommand{\postbb}[3]
{\setlength{\epsfxsize}{#3\hsize}
 \centerline{\epsfbox[#1]{#2}}}

\newcommand{\plb}[2]{{\em Phys. Lett.}              {\bf #1B}, #2 }
\newcommand{\npb}[2]{{\em Nucl. Phys.}              {\bf B#1}, #2 }

\newcommand{\pr }[2]{{\em Phys. Rep.}               {\bf  #1}, #2 }
\newcommand{\prt}[2]{{\em Phys. Rev.}               {\bf D#1}, #2 }
\newcommand{\pru}[2]{{\em Phys. Rev. Lett.}         {\bf  #1}, #2 }

\newcommand{\jp }[2]{{\em J. Phys.}                 {\bf C#1}, #2 }
\newcommand{\sci}[2]{{\em Science}                  {\bf  #1}, #2 }

\newcommand{\epc}[2]{{\em Eur. Phys. J.}            {\bf C#1}, #2 }

\newcommand{\etal}{{\em et al.}}

\newcommand{\ovl}{\overline}
\newcommand{\be}{\begin{equation}}
\newcommand{\ee}{\end{equation}}
\newcommand{\ba}{\begin{array}}
\newcommand{\ea}{\end{array}}

\newcommand{\lsim}{\buildrel < \over {_\sim}}

\begin{document}

\preprint{
\noindent
\hfill
\begin{minipage}[t]{3in}
\begin{flushright}
UPR--T--XXX \\
\vspace*{2cm}
\end{flushright}
\end{minipage}
}

\draft

\title{Indications for an Extra Neutral Gauge Boson in Electroweak 
Precision Data}
\author{Jens Erler and Paul Langacker}
\address{Department of Physics and Astronomy, University of Pennsylvania, 
Philadelphia, PA 19104-6396, USA}

\date{October 1999}

\maketitle

\begin{abstract}
A new analysis of the hadronic peak cross section at LEP~1 implies a small
amount of missing invisible width in $Z$ decays, while the effective weak 
charge in atomic parity violation has been determined recently to 0.6\% 
accuracy, indicating a significantly negative $S$ parameter. As a consequence 
of these two deviations, the data are described well if the presence of 
an additional $Z^\prime$ boson, such as predicted in Grand Unified Theories, 
is assumed. Moreover, the data are now rich enough to study an arbitrary extra 
$Z^\prime$ boson and to determine its couplings in a model independent way. 
An excellent best fit to the data is obtained in this case, suggesting 
the possibility of a family non-universal $Z^\prime$ with properties similar 
to ones predicted in a class of superstring theories. 
\end{abstract}


Neutral gauge structures beyond the photon and the $Z$ boson have long been
considered as one of the best motivated extensions of the Standard Model (SM) 
of electroweak interactions. They are predicted in most Grand Unified Theories 
(GUTs) and appear copiously in superstring theories. Moreover, in the context 
of supersymmetry, they do not spoil the observed approximate gauge coupling 
unification (as predicted by the simplest and most economic of these theories);
in many models (but not the simplest $SO(10)$ ones), the enhanced $U(1)^\prime$
gauge symmetry forbids an elementary bilinear Higgs $\mu$-term, while allowing 
an effective $\mu$ to be generated at the scale of $U(1)^\prime$ breaking 
without introducing cosmological problems~\cite{Cvetic98}. In many 
string-motivated models of radiative breaking, this scale is comparable to 
the electroweak scale (i.e., less than a TeV)~\cite{Cvetic98}, thus providing 
a solution to the $\mu$-problem~\cite{Kim84}, and enhancing the prospects 
to possibly detect it in high energy collisions or to find its effects on 
electroweak precision data. Yet, neither searches at the Tevatron~\cite{Abe97},
nor recent analyses of precision data~\cite{Erler99,Cho98}, yielded evidence 
for the extra neutral $Z^\prime$ boson associated with the $U(1)^\prime$, and 
stringent bounds on its masses and mixings were set.

Following the publication of our recent analysis~\cite{Erler99} 
the contributions of ${\cal O} (\alpha^3)$ to initial state radiation have been
incorporated into the $Z$ lineshape analysis at LEP and the theoretical 
luminosity uncertainty reduced. The new $Z$ lineshape fit~\cite{Mnich99} 
(which includes new experimental input, as well) yields a significantly larger 
and more precise value for the hadronic peak cross section, $\sigma_{\rm had}$,
implying, e.g., for the effective number of massless neutrinos, 
$N_\nu = 2.985 \pm 0.008$, which is $2\sigma$ below the prediction of 3 
standard neutrinos. Therefore, the data favors types of new physics which can 
mimic a negative contribution to the invisible $Z$ decay width. 

The effective weak charge, $Q_W$, in atomic parity violation can be interpreted
as a measurement of the $S$ parameter, with very little sensitivity to 
the other oblique parameters, $T$ and $U$, all of which parametrize new physics
effects due to vector boson self-energies~\cite{Burgers89}. Extra $Z$ bosons 
are not of the pure oblique type, but in the absence of new contributions to 
the $\rho_0$ parameter (or equivalently to $T$) they mimic $S < 0$
if the boson masses satisfy $M_{Z^\prime} > M_Z$~\cite{Amaldi87}. The $Q_W$ 
determination for Cs~\cite{Wood97} has been improved recently to the 0.6\% 
level by measuring the ratio of the off-diagonal hyperfine amplitude (which is 
known precisely~\cite{Bouchiat88}) to the tensor transition 
polarizability~\cite{Bennett99}. As a result, the theory uncertainty was 
reduced significantly and the new $Q_W ({\rm Cs}) = - 72.06 \pm 0.44$ is found 
to be $2.3\sigma$ above the SM prediction.

These two developments necessitate the present update of our analysis in 
Ref.~\cite{Erler99}, partly reversing our previous conclusions. The fit to 
the new precision data improves upon the introduction of certain $Z^\prime$ 
bosons, among them the $Z_\chi$, defined by 
$SO(10) \rightarrow SU(5) \times U(1)_\chi$. Most of the utilized data were 
presented at the International Europhysics Conference on High Energy Physics at
Tampere. For more details and references we refer to the update in 
Ref.~\cite{Erler99A}. Our analysis is based on the FORTRAN routines 
GAPP~\cite{Erler00} including a minor update: for the model-independent fits it
is important to express the hadronic charge asymmetry directly in terms of 
the $Z$ width ratios, $R_q\equiv \Gamma(q\bar{q})/\Gamma(\sum_q q\bar{q})\equiv
\Gamma(q\bar{q})/\Gamma({\rm had})$, and the forward-backward asymmetries,
$$
   Q_{FB} = (\sum\limits_{q = d,s,b} - \sum\limits_{q = u,c}) R_q A_{FB}^{(q)} 
          = 0.04092 \pm 0.00226.
$$

The $U(1)^\prime$ coupling, 
$g^\prime = \sqrt{5/3}\, \sin\theta_W \sqrt{\lambda}\, g_Y$, is defined in 
terms of the hypercharge ($Y$) gauge coupling, $g_Y$, and the weak mixing 
angle, $\theta_W$, such that $\lambda = 1$ if the GUT group breaks directly to 
$SU(3) \times SU(2) \times U(1)_Y \times U(1)^\prime$. $\lambda = 1$ will be 
assumed throughout, but our limits also apply to $\sqrt{\lambda}\, \sin\theta$ 
and ${M_{Z^\prime}\over \sqrt{\lambda}}$ for other values of $\lambda$ (we 
always assume $M_{Z^\prime} \gg M_Z$). Since the $U(1)_\chi$ and $U(1)_Y$ are 
orthogonal at the GUT scale, their kinetic mixing term, 
$F_{\mu\nu}^\prime F^{\mu\nu}$, is small in minimal models. We also assume 
$\rho_0 = 1$, i.e., only Higgs doublets and singlets receive vacuum expectation
values (VEVs). The $Z_\chi$ model has then only $M_{Z^\prime}$ and the 
$Z-Z^\prime$ mixing angle, 
\be
\label{rel1}
   \tan^2 \theta = \frac{M_0^2 - M_Z^2}{M_{Z^\prime}^2 - M_0^2},
\ee
as new parameters, where $M_0$ is the mass of the ordinary $Z$ in the absence 
of mixing. The resulting fit,
$$
\ba{rclrcl}
   M_{Z^\prime} &=& 812^{+339}_{-152} \hbox{ GeV},    \hspace{1pt} &
   \sin\theta   &=& ( - 1.12 \pm 0.80) \times 10^{-3} \hspace{-3pt}, 
   \vspace{3pt} \\
   M_H          &=& 145^{+103}_{-61} \hbox{ GeV}, &
   \alpha_s     &=& 0.1233 \pm 0.0039,
\ea
$$
suggests a TeV scale $Z^\prime$ with small mixing. The lower 
$\chi^2_{\rm min} = 35.35$ for 35 degrees of freedom (d.o.f.) compared with 
the SM ($\chi^2_{\rm min}/{\rm d.o.f.} = 41.88/37$) is mostly due 
to the improvement in the predictions for the atomic parity observables
($\Delta\chi^2 = - 5.67$), but also in $\sigma_{\rm had}$ 
($\Delta\chi^2 = - 2.34$). There is little change in the ratios, 
$R_\ell = {\Gamma ({\rm had}) \over \Gamma (\ell^+ \ell^-)}$ 
($\ell = e$, $\mu$, $\tau$), and the other observables, except that the total 
$Z$ width, $\Gamma_Z$, and the $W$ mass, $M_W$, each contribute about 0.7 more 
to the total $\chi^2$. The improvement in $\sigma_{\rm had}$ is subtle: 
the $Z_\chi$ causes the total and all partial $Z$ decay widths to increase, 
with the exception of the invisible width which decreases by about 0.7~MeV. 
\begin{table}[t]
\centering
\caption{$U(1)^\prime$ charges of a {\bf 27} representation for an arbitrary
$E_6$ boson. Proper $E_6$ normalization is obtained by dividing by 
$2\sqrt{4 a/3 + 5 b/3 + a b + 10/9 + a^2 + 4 b^2}$. The upper half of the Table
with $b=0$ corresponds to a family in $SO(10)$ GUT. $a=0$ describes the $E_6$ 
boson in the absence of kinetic mixing. $D$ ($\ovl{D}$) are exotic 
(anti)-quarks, while the other new states are color singlets.}
\label{E6}
\begin{tabular}{|ccrr|ccrr|}
$\left(\ba{c} \nu \\ e^-     \ea\right)$  & $-1$ &       &       &
$\ba{c} \ovl{\nu} \\ e^+     \ea$ & $\ba{c} +5/3 \\ +1/3 \ea$    &
$\ba{r}       +a  \\  -a     \ea$ & $\ba{r} +2 b \\ + b  \ea$   \\
\hline
$\left(\ba{c}  u  \\  d      \ea\right)$  & +1/3 &       & $+b$  &
$\ba{c} \ovl{u}   \\ \ovl{d} \ea$ & $\ba{c} +1/3 \\ - 1  \ea$    &
$\ba{r}       +a  \\  -a     \ea$ & $\ba{r} + b  \\ {}   \ea$   \\
\hline\hline
$\left(\ba{c}  N  \\ E^-     \ea\right)$  & +2/3 &  $+a$ & $-b$  &
$\ba{c}        D  \\ \ovl{D} \ea$ & $\ba{c} -2/3 \\ +2/3 \ea$    &
$\ba{r}           \\         \ea$ & $\ba{r} -2 b \\ - b  \ea$   \\
\hline
$\left(\ba{c} E^+ \\ \ovl{N} \ea\right)$  &$-2/3$&  $-a$ & $-2b$ & 
$S$ & 0 & & $\ba{r} +3b \ea$ \\
\end{tabular}
\end{table}
Hence, $\Gamma ({\rm had})$ increases more (1.66~MeV) than $\Gamma_Z$ 
(1.15~MeV), implying an overall increase in $\sigma_{\rm had} = 
\frac{12 \pi}{M_Z^2 R_e} \frac{\Gamma ({\rm had})^2}{\Gamma_Z^2}$. As can be 
seen from the fit result, the relatively large increase in $\Gamma ({\rm had})$
is caused by an increased value for $\alpha_s$ 
(SM: $\alpha_s = 0.1192 \pm 0.0028$), in better agreement with expectations
from supersymmetric GUT. At the same time, the $Z_\chi$ does not affect 
the preference for a light Higgs, again in agreement with supersymmetry. 
The value of $\alpha_s$ is, however, slightly higher than the current world
average, $\alpha_s = 0.1182 \pm 0.0013$~\cite{Hinchliffe99}, from 
determinations excluding the $Z$ lineshape. Including this value as 
an additional external constraint yields a result consistent with vanishing 
$Z-Z^\prime$ mixing, $\sin\theta = ( - 0.46 \pm 0.60) \times 10^{-3}$, and 
a smaller decrease in $\chi^2_{\rm min}/{\rm d.o.f.} = 36.85/36$
(SM: $\chi^2_{\rm min}/{\rm d.o.f.} = 42.00/38$). The characteristics of this 
fit are different than the one before: there is a smaller increase in 
$\Gamma (e^+ e^-)$ ($0.03\%$) and there is now a small {\rm decrease} in 
$\Gamma_Z$ ($-0.01\%$), while $\Gamma ({\rm had})$ is virtually unchanged. 
As a result $\sigma_{\rm had}$ is still in better agreement than in the SM, but
at the expense of larger discrepancies in $R_e$ ($1.5\sigma$) and $R_\mu$ 
($1.7\sigma$). Recalling that most of the $\alpha_s$ determinations are
dominated by theoretical uncertainties, it is clearly important to verify 
the very tight (and intrinsically non-Gaussian) constraint on $\alpha_s$, 
which we believe to be asymmetric with a larger positive error.

If the Higgs $U(1)^\prime$ quantum numbers, $Q^\prime_i$, are known, there will
be an extra constraint,
\be
  C = \theta {g_Y\over g^\prime} {M_{Z^\prime}^2\over M_Z^2}
    = - \frac{\sum_i t_{3i} Q^\prime_i |\langle \phi_i \rangle|^2}
             {\sum_i t_{3i}^2          |\langle \phi_i \rangle|^2},
\ee
where the $\langle \phi_i \rangle$ are VEVs, and $t_{3i}$ is the third 
component of isospin. For minimal cases it is given explicitly in Table~III
of Ref.~\cite{Langacker92}. In particular, in $SO(10)$ with light Higgs 
fields from the fundamental {\bf 10} representation, $C = 2/\sqrt{10}$ is 
predicted, and therefore $\theta \sim 5 \times 10^{-3} > 0$. This Higgs 
structure is strongly disfavored by our results.

\begin{table}[t]
\centering
\caption[]{Examples of $Z^\prime$ bosons inspired by $E_6$ symmetry. 
The $Z_\eta$ occurs in Calabi-Yau compactifications of the heterotic string if 
$E_6$ breaks directly to a rank~5 subgroup~\cite{Witten85} via the Hosotani 
mechanism. The $Z_\eta^\ast$ does not couple to leptons and was originally 
introduced to explain a temporarily observed excess of $b$ quark events at 
LEP~\cite{Aguila87}. For the $Z_{LR}$, the values in parentheses are for 
$g_L = g_R$; also assumed for the alternative LR model~\cite{Ma87}, $Z_{ALR}$, 
which changes the fermion assignment, viz.\ $(\nu,e^-)\leftrightarrow (N,E^-)$,
$\ovl{d} \leftrightarrow \ovl{D}$, and $\ovl{\nu} \leftrightarrow S$.
}
\label{E6examples}
\begin{tabular}{|r|rr|rr|}
$Z^\prime$ & $a   $ &   $b$  & $\sin\alpha$ & $\tan\beta$   \\
\hline
$-Z_\chi $ &    0   &    0   &      0       &    0          \\
$ Z_Y    $ & $-5/3$ &    0   &    $-1$      &    0          \\
$ Z_{B-L}$ & $-2/3$ &    0   & $-\sqrt{2/5}$&    0          \\
$-Z_{LR} $ &$>-5/3$ & (1.66)\hspace{40pt}0&$<\sqrt{3/5}$&(0.52)\hspace{40pt}0\\
$ Z_{ALR}$ & $-0.71$& $-0.48$&   $-0.93$    &    $1.47$     \\
$-Z_\psi $ &    0   & $-4/3$ &      0       & $-\infty $    \\
$-Z_\eta $ &    0   & $ 5/3$ &      0       & $-\sqrt{5/3}$ \\
$-Z_\eta^\ast$&$\infty$& $a$ & $\sqrt{8/35}$& $-3/\sqrt{7}$ \\
\end{tabular}
\end{table}

In the $Z_\chi$ model there is no need to introduce exotic fermions beyond
the right-handed neutrino. The question arises what are the most general
$Z^\prime$ couplings consistent with generation-wise cancellation of gauge and 
mixed gauge-gravitational anomalies. The solution can be parametrized by 
the $Z_\chi$ couplings plus a shift proportional to the right-handed isospin
generator, $- 2 a I^3_R$. A non-zero kinetic mixing term~\cite{Aguila95} 
cannot affect the consistency of the theory, and is therefore equivalent to 
this shift. Also any $U(1)^\prime$ originating from an $SO(10)$ GUT can be 
described by $a$. For example, the $Z^\prime$ appearing in left-right (LR) 
models corresponds to $a = g_R^2/g_L^2 \cot^2\theta_W - 5/3$. A fit with $a$ 
yields $a \geq - 0.50$ and,
$$
\ba{rclrcl}
   M_{Z^\prime}     &=& 781^{+362}_{-241} \hbox{ GeV},  \hspace{6pt} &
   \sin\theta       &=& (1.23 \pm 0.89) \times 10^{-3}, \vspace{3pt} \\
   M_H              &=& 165^{+155}_{-91} \hbox{ GeV}, &
   \alpha_s         &=& 0.1239 \pm 0.0045,
\ea
$$
with $\chi^2_{\rm min}/{\rm d.o.f.} = 35.26/34$. The 95\%~CL 
($\Delta\chi^2 = 3.84$) constraint $a \geq - 0.81$ implies for the LR model 
$g_R \geq 0.51\, g_L$, to be compared with the consistency requirement 
$g_R \geq 0.55\, g_L$~\cite{Cvetic92} from the $W^\prime$ sector of these 
models. The LR symmetric case, $g_R = g_L$, yields a lower range for 
$\alpha_s = 0.1210 \pm 0.0032$, and $\chi^2_{\rm min}/{\rm d.o.f.} = 36.05/35$.

\begin{figure}[t]
\postbb{40 220 530 680}{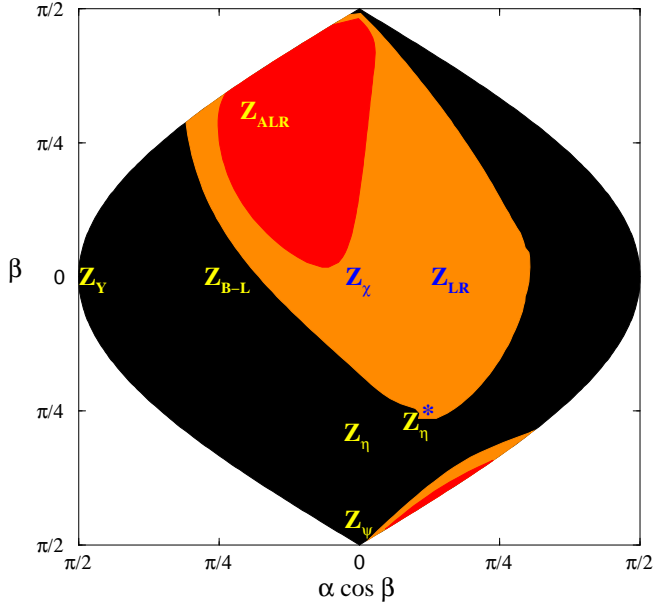}{1.07}
\caption{ Sinusoidal (Sanson-Flamsteed) projection of a hemisphere 
parametrizing arbitrary $E_6$ inspired $Z^\prime$ models. Distances along 
constant $\beta$, the central meridian, and area are preserved.
Shown are the $1\sigma$ and 90\% ($\Delta\chi^2 = 3.04$) contours allowed by 
the data. Note, that while the $Z_\psi$ model (mapped to the poles) does not 
improve the prediction for $Q_W$, it gives a very good fit if some kinetic 
mixing is allowed.}
\label{fig1}
\end{figure}

\begin{figure}
\postbb{40 220 530 680}{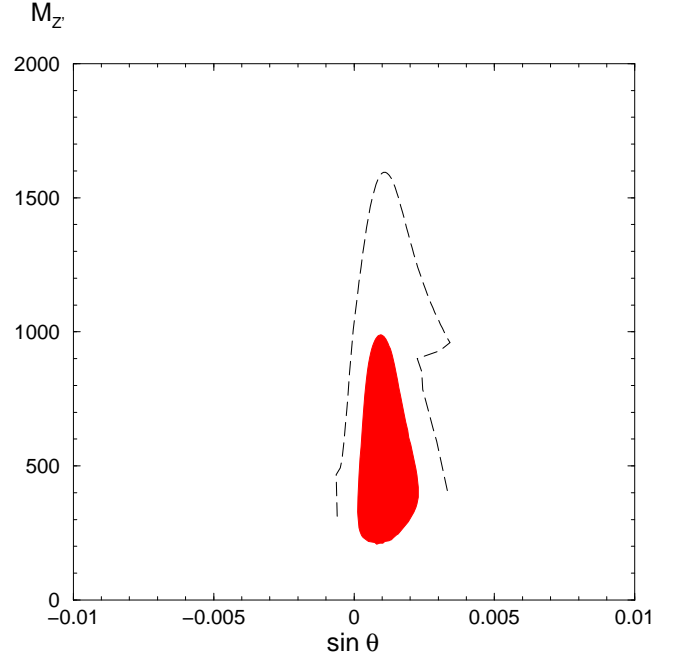}{1.07}
\caption{ $M_{Z^\prime}$ as a function of $\sin\theta$ for an arbitrary $E_6$ 
inspired $Z^\prime$. The contours correspond to the ones in 
Fig.~\ref{fig1}.}
\label{fig2}
\end{figure}

The $E_6$ group contains also a purely axial-vector $Z_\psi$ boson, defined 
through $E_6 \rightarrow SO(10) \times U(1)_\psi$. We define the most general 
$E_6$ boson as a linear combination of the $SO(10)$ boson (parametrized by $a$)
and the $Z_\psi$, via introduction of a second parameter, $b$. Alternatively, 
\be
  Z^\prime \sim - \cos\alpha \cos\beta\, Z_\chi 
                + \sin\alpha \cos\beta\, Z_Y - \sin\beta\, Z_\psi,
\ee
can be expanded in terms of an orthonormal basis, with
\be
\ba{rcl}
  a &=& \frac{ - \sin\alpha}{{3\over 5}\sin\alpha - \sqrt{27\over 50}\cos\alpha
               - \sqrt{1\over 10} \tan\beta}, \vspace{5pt} \\
  b &=& - \sqrt{8\over 45} {\tan\beta \over \sin\alpha} a.
\ea
\ee
The couplings of the states in the {\bf 27} are listed in Table~\ref{E6}. 
The angles $\alpha$ and $\beta$ parametrize the surface of a sphere where
opposite points correspond to the same model with the opposite sign of 
$\theta$. Therefore, we can restrict ourselves to one hemisphere defined by
$-\pi/2 \leq \alpha,\beta < \pi/2$. The values of the parameter sets $(a,b)$ 
and $(\alpha,\beta)$ are shown for popular cases in Table~\ref{E6examples}. 
A fit with both $a$ and $b$ allowed yields 
$\chi^2_{\rm min}/{\rm d.o.f.} = 34.21/33$, and
$$
   M_H      = 131^{+131}_{-64} \hbox{ GeV}, \hspace{30pt}
   \alpha_s = 0.1232_{-0.0046}^{+0.0050}.
$$
The $1\sigma$ contour and the 90\%~CL allowed region obtained by integrating 
over the hemisphere (giving each point equal prior probability) is shown in 
Fig.~\ref{fig1}. The corresponding contours in $M_{Z^\prime}$ vs. 
$\sin\theta$ are shown in Fig.~\ref{fig2}. The $1\sigma$ contour in 
Fig.~\ref{fig1} shows that $a$ is consistent with zero, while $b < 0$. If we 
repeat this fit with $a=0$ fixed (no kinetic mixing), we find 
$\alpha_s = 0.1216 \pm 0.0036$ closer to the outside constraint. Including it 
in the fit yields 
$$
\ba{rclrcl}
   M_{Z^\prime}     &=& 287^{+673}_{-101} \hbox{ GeV},  \hspace{6pt} &
   \sin\theta       &=& (0.36 \pm 0.63) \times 10^{-3}, \vspace{3pt} \\
   M_H              &=& 101^{+57}_{-39} \hbox{ GeV}, &
   \alpha_s         &=& 0.1186 \pm 0.0012 \vspace{3pt}, \\
   b                &=& - 1.06_{-0.16}^{+0.91}, & & & \hspace{-30pt}
   \chi^2_{\rm min}/{\rm d.o.f.} = 35.58/35.
\ea
$$
This concludes our discussion of $E_6$ bosons, which constitute a very well 
motivated class of cases, both in GUT and superstring contexts, and many of 
which significantly improve the fit to the precision data. 

A family-universal $Z^\prime$ with otherwise arbitrary couplings is described 
by the five charges of the SM multiplets (cf.\ Table~\ref{E6}), and 
the mixing angle, $\theta$. $M_{Z^\prime}$ cannot be determined independently 
using the precision data, and will be fixed to 1~TeV in the following (for 
other values the charges and mixing scale as $M_{Z^\prime}$ and 
$M_{Z^\prime}^{-1}$, respectively). We also employ the $\alpha_s$ 
constraint~\cite{Hinchliffe99}, as otherwise the strong coupling is driven to 
very low and clearly excluded values. The resulting fit is good, 
$\chi^2_{\rm min}/{\rm d.o.f.} = 31.73/32$, and yields the couplings,
$$
\ba{rclrcl}
   (\nu,e)_L  &=& -0.24_{-0.23}^{+0.12}, &
   e_R        &=& -0.21_{-0.14}^{+0.12}, \vspace{3pt} \\
   (u,d)_L    &=& -0.16_{-0.92}^{+0.30}, &
   d_R        &=&  1.32_{-0.51}^{+1.35}, \vspace{3pt} \\
   \sin\theta &=& (4.04^{+0.69}_{-2.72}) \times 10^{-3}, \hspace{4pt} &
   u_R        &=&  0.05_{-0.67}^{+1.22}.
\ea
$$
Stronger mixing is allowed than in the $E_6$ cases, and we find the 95\% CL 
bound,
\be
  | \sin\theta \, {M_{Z^\prime}\over 1~{\rm TeV}} | \leq 5.31 \times 10^{-3},
\label{thetabound}
\ee
mainly from $M_W$. The leptonic couplings, $e_L \approx e_R$, as well as $d_R$
and $\theta$, all are significantly different from zero and strongly 
correlated. In contrast to the $E_6$ inspired cases, the tendency for a light 
Higgs is lost; instead we find $M_H \geq 210$ GeV for the $1\sigma$ ranges 
above. Another family-universal possibility is the sequential $Z_{\rm SM}$, 
with the same couplings as the $Z$. For example, this could be due to 
the excited $Z$ states in models with extra TeV-scale dimensions. 
However, the $Z_{\rm SM}$ would yield a {\em larger} $|Q_W ({\rm Cs})|$, 
increasing the discrepancy with experiment.

In general, the extra $Z$ boson may not couple in a universal way. There are,
however, strong constraints from flavor changing neutral current processes
specifically limiting non-universality between the first two generations.
On the other hand, the third generation is less constrained. We therefore treat
the first two generations as universal, while allowing arbitrary couplings to
the third. Since the $U(1)^\prime$ charge of the top quark does not enter (at
tree level), there are four new parameters compared to the universal fit;
we find,
$$
\ba{rclrcl}
   (\nu,e)_L    &=& -0.32_{-0.69}^{+0.13}, \hspace{4pt} & 
   (\nu,\tau)_L &=& -0.24_{-0.57}^{+0.78}, \vspace{3pt} \\ 
   e_R          &=& -0.31_{-0.60}^{+0.13}, & 
   \tau_R       &=&  0.03_{-0.25}^{+1.92}, \vspace{3pt} \\
   (u,d)_L      &=& -0.52_{-4.78}^{+1.11}, & 
   (t,b)_L      &=&  1.32_{-0.72}^{+9.10}, \vspace{3pt} \\
   d_R          &=&  1.72_{-1.25}^{+5.61}, & 
   b_R          &=&  8.48_{-3.16}^{+52.4}, \vspace{3pt} \\
   u_R          &=&  0.38_{-1.18}^{+4.85}, &
   \sin\theta   &=& (4.06^{+0.69}_{-3.51})\times 10^{-3}, 
\ea
$$
and an excellent $\chi^2_{\rm min}/{\rm d.o.f.} = 25.33/28$, while 
the bound~(\ref{thetabound}) is virtually unchanged. Six of the ten quantities 
are significantly different from zero, and all couplings are of similar 
magnitude and ${\cal O}(1)$ (for a TeV scale $Z^\prime$), except for $b_R$ 
which appears surprisingly large (driven by the experimental anomaly in 
the asymmetry parameter $A_b$ at LEP and SLC). It is amusing, however, that 
$b_R/u_R = 22.62$ is consistent with the prediction (22.67) of a closely 
investigated superstring model~\cite{Chaudhuri95,Cleaver98} (based on 
the fermionic construction of the heterotic string). Of course, some other 
predictions fail, but it is important to keep in mind that unlike in GUT 
models, superstrings often predict family non-universal $Z^\prime$ bosons with 
highly unconventional coupling ratios. The $\chi^2_{\rm min}$ is lower by 
$-16.67$ for 10 d.o.f.\ as compared to the SM, which has a probability of 8.2\%
to be due to a fluctuation. 

We therefore conclude, that the data are generally better described if 
the presence of an extra $Z^\prime$ boson is assumed. There are many 
realizations of such a possibility, ranging from the standard and much more 
restrictive GUT scenarios to the more exotic bosons as predicted in some string
models. Of course, a light $Z^\prime$ precludes a heavy Majorana neutrino, as 
is invoked in seesaw models and some models of baryogenesis, except in 
the cases that the neutrino (e.g., $\ovl{\nu}$ or $S$ in Table~\ref{E6}) 
carries no $U(1)^\prime$ charge. A $Z^\prime$ with mass $\lsim 1$~TeV may be 
directly observable by its leptonic decays in Run II at the Tevatron, and 
certainly at the LHC. For this mass range, a variety of complementary 
diagnostics of its couplings will be possible at the LHC, a future lepton 
collider, as well as from the precision data~\cite{Cvetic98,Cvetic95}.

\centerline{\bf Acknowledgements:}
This work was supported in part by the US Department of Energy grant 
EY--76--02--3071.

\end{document}